\documentclass[preprint,5p,twocolumn]{elsarticle}

\usepackage[font=small,labelfont=bf]{caption}
\usepackage{url}

\usepackage{booktabs} 
\usepackage{amsmath,amssymb,amsfonts}
\usepackage{algorithmic}
\usepackage{graphicx}
\usepackage{xspace}
\usepackage{textcomp}
\usepackage{subcaption}
\usepackage{siunitx}
\usepackage{booktabs}
\usepackage{multirow}
\usepackage{subfiles}
\usepackage{xspace}
\usepackage{tablefootnote}

\usepackage[numbers]{natbib} 

\usepackage{mathtools}
\usepackage{soul}
\usepackage{comment}

\begin{document}
\begin{frontmatter}

\title{Experimenting with Adaptive Bitrate Algorithms for Virtual Reality Streaming over Wi-Fi}

\journal{arXiv}

\author{Ferran Maura\corref{cor1}}
\ead{ferranjosep.maura@upf.edu}
\author{Miguel Casasnovas\corref{cor1}}
\ead{miguel.casasnovas@upf.edu}
\author{Boris Bellalta\corref{}}
\ead{boris.bellalta@upf.edu}
\address{Wireless Networking Research Group, Universitat Pompeu Fabra, Carrer de Roc Boronat 138, 08018 Barcelona, Spain}
\cortext[cor1]{Equal contribution}


    



\begin{abstract} 
Interactive Virtual Reality (VR) streaming over Wi-Fi networks encounters significant challenges due to bandwidth fluctuations caused by channel contention and user mobility. Adaptive BitRate (ABR) algorithms dynamically adjust the video encoding bitrate based on the available network capacity, aiming to maximize image quality while mitigating congestion and preserving the user's Quality of Experience (QoE). In this paper, we experiment with ABR algorithms for VR streaming using Air Light VR (ALVR), an open-source VR streaming solution. We extend ALVR with a comprehensive set of metrics that provide a robust characterization of the network's state, enabling more informed bitrate adjustments. To demonstrate the utility of these performance indicators, we develop and test the Network-aware Step-wise ABR algorithm for VR streaming (NeSt-VR). 
Results validate the accuracy of the newly implemented network performance metrics and demonstrate NeSt-VR's video bitrate adaptation capabilities.
 
\end{abstract}





\end{frontmatter}


\section{Introduction}

Cloud/Edge-based eXtended Reality (XR)\footnote{XR encompasses Virtual Reality (VR), Augmented Reality (AR), and Mixed Reality (MR)} streaming applications are becoming increasingly popular, offering new opportunities in several fields such as healthcare, industry, education, and gaming \cite{minopoulos2022opportunities}. However, streaming XR content poses significant challenges to current and future communication networks due to its stringent throughput, latency, and reliability requirements~\cite{akyildiz2022wireless}. For instance, for VR streaming, the Wi-Fi Alliance recommends video bitrates in the range between 100 and 200~Mbps, with RTT values below 7~ms~\cite{wifialliance_VR_reqs}.



Wi-Fi is anticipated to become the main technology for XR streaming \cite{giordano2023will}, given the predominance of indoor XR use cases and its ability to provide low-latency, high-reliability, and energy-efficient connections with multi-gigabit speeds. 
Indeed, Wi-Fi serves as the last hop in applications such as Steam Link\footnote{\url{$https://store.steampowered.com/app/353380/Steam_Link/$}}, Quest Air Link, Virtual Desktop\footnote{\url{https://www.vrdesktop.net/}}, and Air Light VR (ALVR)\footnote{\url{https://github.com/alvr-org/ALVR/}}. These interactive VR streaming applications offload computationally intensive tasks, such as rendering, from Head-Mounted Displays (HMDs) to cloud or edge computing servers, addressing the HMDs’ limited battery life and processing power while facilitating high-quality graphics and the use of lighter, more comfortable devices. 

A significant challenge in deploying interactive VR streaming solutions is the variability in the network’s available bandwidth. During network congestion intervals ---manifested through packet losses and high latency--- the network may struggle to sustain video traffic. This can result in lost and delayed frames, leading to stuttering and visual artifacts that significantly degrade the user's Quality of Experience (QoE). Thus, Adaptive BitRate (ABR) algorithms are required. These algorithms dynamically adjust the video encoding bitrate based on the network state, optimizing bitrate selection to enhance image quality. 

ABR implementations vary widely across the state-of-the-art \cite{ABR_survey}. Given the unique demands of interactive VR streaming, such as minimal jitter buffer to reduce latency and high throughput for optimal image quality, traditional ABR buffer-based algorithms such as \cite{buffer-based-netflix, BOLA} ---that are nearly optimal for pre-buffered video--- are ineffective in this context. Therefore, ABR solutions based on network capacity estimation are a feasible choice for VR streaming. 
For instance, EVeREst, a control heuristic designed to ensure timely delivery of video frames (VFs) without exceeding the available network capacity is introduced in \cite{liubogoshchev2021everest, korneev2024model}. A deep reinforcement learning policy to optimize latency, network parameters, and visual quality indicators ---such as Peak Signal-to-Noise Ratio (PSNR)--- for 5G VR streaming is proposed in \cite{ALVR_5G_DQN}. Lastly, a scheme that considers the visual complexity of VFs and applies a mechanism for bitrate control inspired by \cite{GCC-webrtc} is presented in \cite{alhilal2024fovoptix}. Lastly, a fuzzy logic algorithm to optimize TCP transport for enhanced QoE in VR is introduced in \cite{vergados2023adaptive}.

In this paper, we investigate ABR algorithms tailored for VR streaming using ALVR, an open-source VR streaming solution. 
We enhance ALVR with an extensive set of performance metrics that provide a detailed characterization of the network's state, facilitating more informed bitrate adjustments. To demonstrate the utility of these metrics, we develop the Network-aware Step-wise ABR algorithm for VR streaming (NeSt-VR), testing it under scenarios of network capacity fluctuations and user mobility.
In particular, the contributions of this paper are:
\begin{enumerate}
    \item Extending ALVR's source code with $i)$~a set of performance metrics to characterize the network state, and $ii$)~a mechanism to receive timely feedback from the HMD at the server upon reception of a VF. 
    \item Presenting a methodology for validating network-related metrics, leveraging Wireshark traffic traces.
    \item Proposing the Network-aware Step-wise ABR algorithm for VR (NeSt-VR), designed to `conservatively' react to network disruptions and capacity changes using several of our implemented metrics.
    \item Conducting an experimental evaluation of both NeSt-VR and ALVR's native ABR solution to demonstrate their operational differences and superior performance over a Constant BitRate (CBR) approach in dynamic network environments. 
\end{enumerate}

\section{Air Light VR (ALVR)}

\subsection{Overview}
\begin{figure}[t!!!]
    \centering
    \includegraphics[width=\columnwidth]{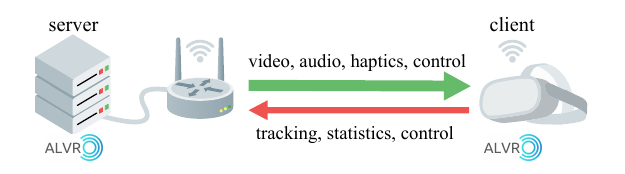}
    \caption{ALVR streaming process.}
    \label{Fig:VRStreamingPipeline}
\end{figure}

\sloppy
ALVR is an open-source project that interfaces SteamVR\footnote{\url{https://store.steampowered.com/app/250820/SteamVR}} to enable wireless streaming of VR content from a server to an untethered standalone HMD ---such as the Meta Quest~2--- over Wi-Fi. As depicted in Fig.~\ref{Fig:VRStreamingPipeline}, the server ---acting as the streamer--- sends video, audio, control data, and haptic feedback in the downlink (DL) to the HMD. On the other hand, the HMD ---functioning as the client--- transmits tracking, control, and statistical data in the uplink (UL) to the server.
\fussy

In particular, the client gathers tracking data from its sensors and controllers, sending it to the server. On the server, this data is fed into SteamVR for pose prediction, the VR game logic is executed, and the graphical layers are rendered. 
These layers are combined into a VF that is encoded and fragmented into multiple packets\footnote{Each packet carries an application-specific prefix that includes: a frame index, a packet index within the frame, the number of packets composing the frame, the frame size in bytes, and an stream identifier} that are transmitted to the client.  
Upon reception, the client reassembles, decodes, and processes the VF, submitting it to the VR runtime for rendering and display. At that moment, the client also sends statistical data associated with the VF to the server for logging.

\begin{figure}[t!!!]
     \centering
     \includegraphics[width=\columnwidth]{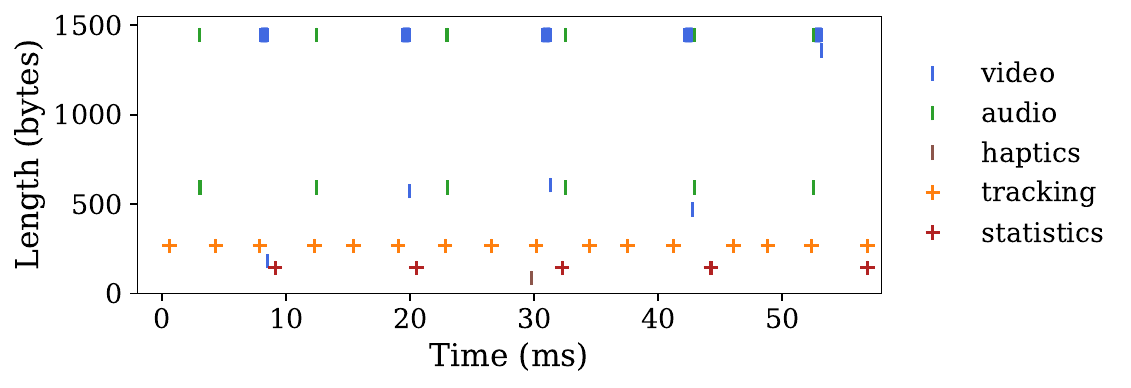}
     \caption{ALVR v20.6.0 traffic using CBR at 100 Mbps, 90 fps, UDP, and SteamVR Home. Derived from parsed Wireshark traffic traces. Packet length includes transport headers.}
     \label{Fig:ALVR_traffic}
     \end{figure}

As illustrated in Fig.~\ref{Fig:ALVR_traffic}, in the DL, VFs are transmitted in single bursts every $1/\text{fps}$ (i.e., every 11 ms at 90 fps) and audio data is transmitted every 10 ms, both including 1446-byte packets\footnote{The maximum packet payload size in ALVR is configurable and defaults to 1400 bytes} and a remainder of variable size. Conversely, haptic feedback is sent occasionally using 88-byte packets, depending on user actions and application-specific events. On the other hand, in the UL, tracking data is transmitted at a frequency of three updates per frame using 207-byte packets, while statistical data is sent once ---upon submission of a VF to the VR runtime--- using a 144-byte packet. Note that the aforementioned packet sizes include transport headers.

\subsection{ABR algorithm}

ALVR implements both CBR and ABR for video transmission. ALVR's ABR algorithm dynamically adjusts the encoder’s target bitrate once per second. At each adjustment step, it uses the moving average of ALVR's capacity estimate ($C_{\text{ALVR}}$) as the initial bitrate, then scaled by a configurable multiplier (default 0.9) to ensure the stream’s bitrate remains under the network’s capacity. Note that $C_{\text{ALVR}}=\frac{L}{d_{\text{ntw}}}$,
where $L$ represents the payload size of a VF and $d_{\text{ntw}}$ denotes the network delay\footnote{Network delay represents the latency in the motion-to-photon pipeline due to data transmission, including the time required for a tracking packet to travel in the UL and a complete VF ---fragmented into multiple packets--- to travel in the DL} associated with the VF transmission. 

This initial bitrate is then constrained ---within the configured maximum and minimum bitrate limits--- in response to excessive encoder, decoder, and network delays, surpassing configurable static thresholds: $0.9 \cdot (1/\text{fps}_t)$~ms, $30$~ms, and $8$~ms by default, respectively, where $\text{fps}_t$ denotes the targeted frame rate. In particular, these upper limits are imposed based on the latencies' deviation from their thresholds, scaling the initial bitrate proportionally.

\section{Experimental Setup}

\subsection{Equipment}

Our experimental setup includes a high-performance VR-Ready Personal Computer (PC) used as a server, a basic PC serving as a network emulator (netem), a gaming-centric Wi-Fi Access Point (AP), and a Meta Quest~2 HMD.  Details can be found in Tbl.~\ref{tab:experiment_conditions}.

\begin{table}[t]
    \centering
    \caption{Equipment details.}
    \small
    \begin{tabular}{@{}p{2.5cm} p{2cm} p{3cm}@{}}
\toprule
    \textbf{Equipment} & \textbf{Specification} & \textbf{Details} \\
\midrule
    \multirow{4}{*}{\textbf{VR-Ready PC}} & OS & Windows 10 x64 \\
    & GPU & NVIDIA GeForce RTX 3080, 10 GB \\
    & CPU & i5-12600KF \\ 
    & NIC & Realtek Gaming 2.5GbE \\
\midrule
    \multirow{5}{*}{\textbf{Netem PC}} & Model & Dell Optiplex 3000 \\
    & OS & Ubuntu Desktop 22.04.3 \\
    & GPU & Intel UHD Graphics 770 \\
    & CPU & i5-12500 \\ 
    & NIC (default) & Realtek RTL8111 1GbE \\ 
    & NIC (added) & Realtek Gaming 2.5GbE \\ 
\midrule
    \multirow{3}{*}{\textbf{AP}} & Model & ASUS ROG Rapture
    GT-AXE11000 \\
    & Firmware & 3.0.0.4.388\_22525 \\
    & Standard & 802.11ax \\
\midrule
    \textbf{HMD} & Model & Meta Quest 2 \\
\bottomrule
    \end{tabular}
    \label{tab:experiment_conditions}
\end{table}

\begin{figure}[t!!!]
    \centering
    \begin{subfigure}{\columnwidth}
        \centering
        \includegraphics[width=0.95\linewidth]{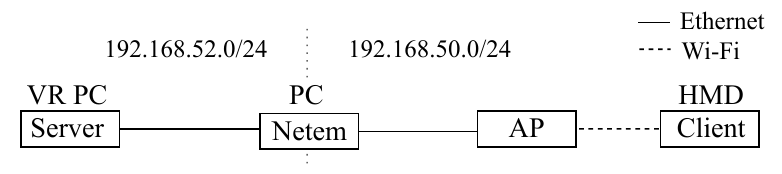}
        \caption{Network topology.}
    \end{subfigure}
    
    \vspace{4mm}
    \begin{subfigure}{0.65\columnwidth}
        \centering
        \includegraphics[width=\linewidth]{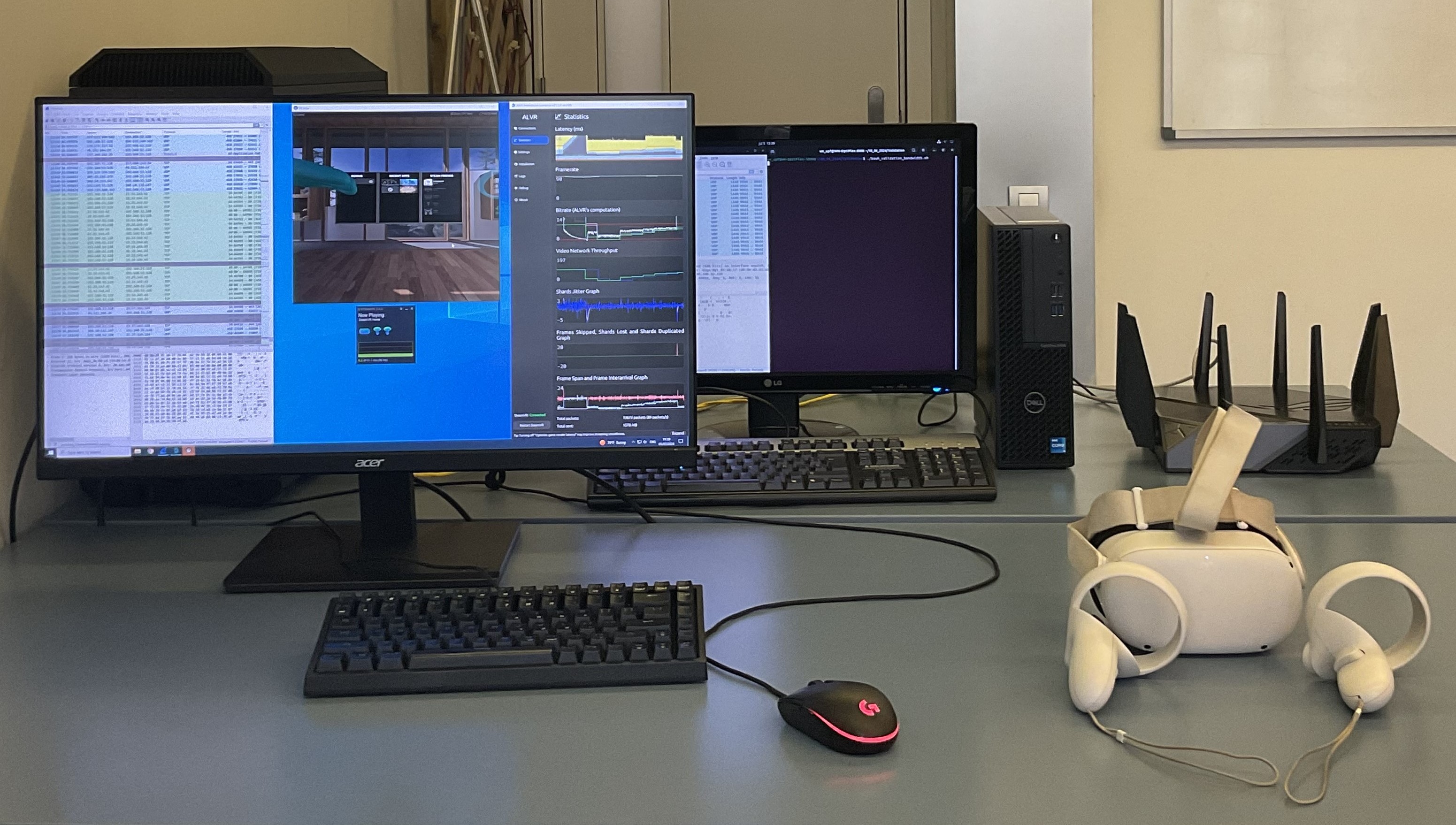}
        \caption{VR-Ready PC, Netem PC, AP, and HMD.}
    \end{subfigure}
    \caption{Testbed components.}
    \label{Fig:testbed}
\end{figure}

As illustrated in Fig.~\ref{Fig:testbed}, the PCs and AP are connected via a 1~Gbps Ethernet cable. Conversely, the HMD is connected over Wi-Fi to the AP. The VR-ready PC, running Windows, serves as the streamer, while the Meta Quest~2 HMD serves as the client, both executing the binaries produced by our ALVR~v20.6.0 fork\footnote{\url{https://github.com/wn-upf/ALVR_ABR_UPF}}. The AP and HMD use Wi-Fi~6 at 5~GHz with an 80~MHz channel bandwidth and no multi-user features enabled. The netem, running Linux, is placed between the VR-ready PC and the AP and uses Linux's traffic control~(tc)\footnote{\url{https://man7.org/linux/man-pages/man8/tc.8.html}} to emulate several network conditions such as packet loss, jitter, and limited bandwidth in the DL. 

\subsection{Methodology}

Our tests in Sections~\ref{Sec:validation} and \ref{Sec:abr_evalu} span 70 and 120 seconds, respectively. Section~\ref{Sec:validation} tests are conducted without a congestion control mechanism, using a CBR of 100~Mbps. 
On the other hand, Section~\ref{Sec:abr_evalu} compares distinct bitrate management approaches, including CBR (100~Mbps), ALVR's ABR algorithm (10 to 100~Mbps), and NeSt-VR (10 to 100~Mbps). All tests stream SteamVR Home at 90~fps, using our ALVR~v20.6.0 fork default settings, including HEVC and UDP. 


\subsection{Datasets}
Our dataset, publicly available on Zenodo\footnote{\url{https://doi.org/10.5281/zenodo.12723990}\label{fn:zenodo}}, includes ALVR session logs (.json) for each test, encompassing our metrics. Additionally, for each test in Section \ref{Sec:validation}, it includes \textit{tshark}-processed traffic traces (.tsv) collected using Wireshark~v4.0.3 at both the server and the netem's Ethernet interface to the AP. Note that there is no packet capture application available for a Meta Quest 2. Thus, the server's interface captures the generated video packet stream and the UL packets received, while the netem's interface captures the stream after the emulation of network effects. 

\section{WN-ALVR extensions}


We have created an ALVR fork that incorporates several network performance metrics, aiding the decision-making process of any ABR algorithm. For instance, metrics like the arrival span and inter-arrival time of VFs can be used to replicate VR-specific ABR algorithms such as EVeREst~\cite{liubogoshchev2021everest, korneev2024model}. 

Modifications have been made throughout ALVR's pipeline to gather client-side statistics ---essential for obtaining the network performance metrics of interest--- and relay them to the server for their logging and utilization in our implemented ABR algorithm, NeSt-VR. Notably, since ALVR may drop decoded VFs prior to visualization if the queue of decoded frames overflows, thereby preventing the transmission of their associated native statistics packets, our client-side statistics are encapsulated in an additional 56-byte UL packet. This packet is sent over TCP in the UL immediately after the client receives a VF. Hence, for a given VF, the feedback is received at the server significantly sooner than using ALVR’s native statistics packet, which is delayed until the VF is ready for display. 

\subsection{Characterizing Network Performance}\label{sec:network-metrics}

Here, we introduce the new performance metrics integrated into ALVR to gather information about the network's performance.

\subsubsection{Time-related metrics}
    \begin{itemize}
        \item \textbf{Client-side frame span}: time interval between the reception of the first packet to the reception of the last packet of a VF. 
        %
        \item \textbf{Frame inter-arrival time}: time interval between the reception of two consecutive complete VFs. It is indicative of the network’s consistency in delivering VFs.
        \item \textbf{Video Frame Round-Trip Time (VF-RTT)}: time it takes for a VF to travel from the server to the client and for our supplementary UL packet ---promptly sent upon the complete reception of the VF--- to reach the server.
    \end{itemize}

\subsubsection{Reliability metrics}
    \begin{itemize}
        \item \textbf{Packet loss}: number of packets lost in the interval between two VF receptions. It leverages the sequence number\footnote{The sequence number serves as a unique identifier for each packet within the global sequence of transmitted packets} of each video packet received to estimate the number of packets sent during the interval. 
    \end{itemize}
\subsubsection{Data rate metrics}
    \begin{itemize}
        \item \textbf{Instantaneous video network throughput}: 
        rate at which video data is received by the client, measured in the interval between two VFs receptions.
        It reflects the network's capability to sustain the desired video quality according to the encoder's target bitrate. 
        \item \textbf{Peak network throughput}: it is computed as the ratio between the VF's size and its client-side frame span. It serves as an estimate of the network capacity since ALVR sends each VF in a single burst.
    \end{itemize}
    
\subsubsection{Network Stability metrics}

    \begin{itemize}
        \item \textbf{VF jitter}: variation in VF time deliveries, providing insight into the smoothness of video playback. It is computed as the sample standard deviation of frame inter-arrival times.
        %
        \item \textbf{Video packet jitter}: variability in video packet arrival times, as defined in RFC 3550 \cite{rfc3550}, providing insight into the consistency of packet delivery.
        \item \textbf{Filtered one-way delay gradient (FOWD)}\footnote{FOWD is used in delay-based congestion control algorithms such as Google Congestion Control (GCC) \cite{GCC-webrtc}}: indicates the rate of change in one-way delay\footnote{To measure one-way delay we have incorporated the transmission time of each packet into ALVR’s application prefix} between two consecutive VFs, smoothed using a Kalman filter as described in \cite{GCC-webrtc} and a state noise variance of $10^{-7}$.
    \end{itemize}

\subsection{NeSt-VR}\label{Sec:heuristic}

NeSt-VR is a balanced ABR algorithm designed to dynamically adjust the video bitrate in response to network congestion intervals during the streaming of VR content over Wi-Fi. NeSt-VR operates every $\tau$ seconds, progressively adjusting the target bitrate ($B_{\text{v}}$) ---initially set to $ B_{0}$~Mbps--- in $\beta$~Mbps steps to avoid significant video quality shifts that may disrupt the user’s QoE. 

\begin{figure}[t!!!]
    \centering
    \includegraphics[width=0.99\columnwidth]{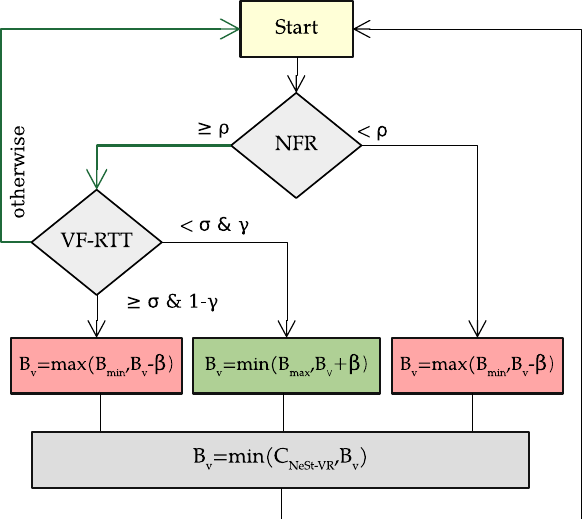}
    \caption{NeSt-VR algorithm diagram. 
    }
    \label{Fig:SimpleHeur}
\end{figure}

Given network congestion introduces delays and losses that significantly disrupt the reception of VFs, NeSt-VR uses the Network Frame Ratio (NFR) and VF-RTT as inputs. These metrics are averaged over an $n$-sample sliding window (~$\overline{\;\centerdot\;}$~) to avoid overreacting to severe but sporadic congestion episodes. 

NFR provides insights into the network’s reliability and consistency in delivering VFs: $\overline{\rm NFR} =\overline{\text{fps}_{\rm rx}}/{\overline{\text{fps}_{\rm tx}}}$. Here, $\text{fps}_{\rm rx}$ denotes the frame delivery rate\footnote{Frame delivery rate denotes the frames per second successfully and timely received at the client after network transmission}, such that $\overline{\text{fps}_{\rm rx}} = 1/\overline{\Delta_{\rm rx}}$, and $\text{fps}_{\rm tx}$ denotes the frame transmission rate, such that $\overline{\text{fps}_{\rm tx}} = 1/\overline{\Delta_{\rm tx}}$. Here, $\Delta_{\rm rx}$\footnote{Note that $\Delta_{\rm rx}$ is equivalent to the frame inter-arrival time} and $\Delta_{\rm tx}$ correspond to the intervals between consecutive VFs receptions and transmissions, respectively.

As illustrated in Fig. \ref{Fig:SimpleHeur}, NeSt-VR applies a hierarchical decision-making process:
\begin{itemize}
    \item If NFR is below the threshold $\rho$, the target bitrate is reduced to alleviate the network strain, minimizing packet losses and enhancing the frame delivery rate.
    \item If NFR exceeds $\rho$ but VF-RTT is below its threshold $\sigma$, given by $\sigma=\varsigma/\overline{\Delta_{tx}}$, the network may be capable of sustaining a higher bitrate. Thus, with probability $\gamma$\footnote{$\gamma$ serves as an exploration parameter to assess whether higher bitrates can be sustained without compromising the user’s QoE. It also moderates the frequency of bitrate adjustments influenced by high VF-RTTs, thereby assigning less significance to VF-RTT compared to NFR in bitrate adaptation decisions}, the bitrate is increased; otherwise, it remains consistent.
    \item If both NFR and VF-RTT surpass $\rho$ and $\sigma$, respectively, it indicates adequate VF delivery rate but significant delay in VF arrivals, impacting motion-to-photon latency. Thus, with a probability of $1\text{-}\gamma$, the bitrate ---and consequently, also the number of packets per VF--- is reduced. This decrease in bitrate significantly reduces DL delays given that the queuing at the AP is shortened. 
\end{itemize}

\setlength{\tabcolsep}{4pt} 
\begin{table}[ht]
    \centering
    \caption{NeSt-VR parameters.}
    \small
    \begin{tabular}{@{}ll|ll@{}}
    \toprule
    Adjustment period & $\tau$ & Step size & $\beta$\\
    Sliding window size & $n$ & Est. Capacity scaling factor & $m$\\
    min Bitrate & $B_{\min}$ & VF-RTT Exploration Prob. & $\gamma$ \\
    max Bitrate & $B_{\max}$ & NFR thresh. & $\rho$ \\
    initial Bitrate & $B_{0}$ & VF-RTT thresh. scaling factor & $\varsigma$  \\
  
    \bottomrule 
    \end{tabular}
    \label{tab:heur-values}
\end{table}

Finally, the target bitrate is constrained within the configured maximum and minimum bitrate limits ($B_{\max}$ and $B_{\min}$) and is further upper bounded by $m \cdot C_{\text{NeSt-VR}}$ ---with $m \leq 1$--- to  ensure the bitrate remains under the network’s capacity.
Here, $C_{\text{NeSt-VR}}$ denotes NeSt-VR’s estimated network capacity, computed as the average peak network throughput over a sliding window of $n$ samples.
Tbl.~\ref{tab:heur-values} outlines NeSt-VR's parameters.
\section{Metrics validation} \label{Sec:validation}

The metrics introduced in Section~\ref{sec:network-metrics} have been independently compared against their measured counterparts derived from parsed Wireshark traffic traces\footnote{Netem traffic traces were employed to validate frame span, frame inter-arrival, and network throughput, while server traces were used to validate VF-RTT. Both netem and server traces contributed to validate packet loss and FOWD} to assess their reliability and representativeness. In particular, the application prefix within each packet's payload has been parsed to identify VFs and the metrics have been computed independently but in an analogous manner to ALVR ---logging metrics only for completely and timely received VFs.

Several tests emulating individual network effects have been conducted: $i)$ limited bandwidth; $ii)$ packet loss; $iii)$ duplicated packets; and $iv)$ packet-level jitter. Each test comprised three 10-second intervals of emulated effects at varying intensities, interspersed with 10-second intervals in normal conditions. Note that using netem’s Wireshark traces leads to optimistic measurements, as it does not account for Wi-Fi transmission delays.\footnote{Discrepancies between results from netem Wireshark traces and ALVR logs can be used to isolate the Wi-Fi impact on the system's network performance} 
Thus, to minimize the Wi-Fi’s hop impact on packet timings, the client was positioned close to the AP ($<$1~m) with a Received Signal Strength Indicator (RSSI) of -40~dBm.
The specific emulation effects and their levels at each interval are detailed in Tbl.~\ref{table:emulation_tests}. Due to space constraints, in this paper, we present only our results from test $i)$. Validation results for other scenarios are available in Zenodo\footref{fn:zenodo}.

\begin{table}[t!]
\centering
\caption{Emulated network effects and their intensities.
}
\small
\begin{tabular}{@{}l l  l  l  l @{}}
\toprule
&& \textbf{Interval 1} & \textbf{Interval 2} & \textbf{Interval 3} \\
\midrule
$1$ &Lim. bandwidth & 100 Mbps & 95 Mbps & 90 Mbps \\
$2$ &Packet loss & 0.5\% & 1\% & 2\% \\
$3$ &Duplicated pkts & 0.5\% & 1\% & 2\% \\
$4$ &Packet jitter & 0-6 ms & 0-10 ms & 0-20 ms \\
\bottomrule
\end{tabular}

\label{table:emulation_tests}
\end{table}

\begin{figure*}[t] %
  \centering
  \begin{subfigure}[b]{1.0\linewidth} 
    \includegraphics[width=\linewidth]{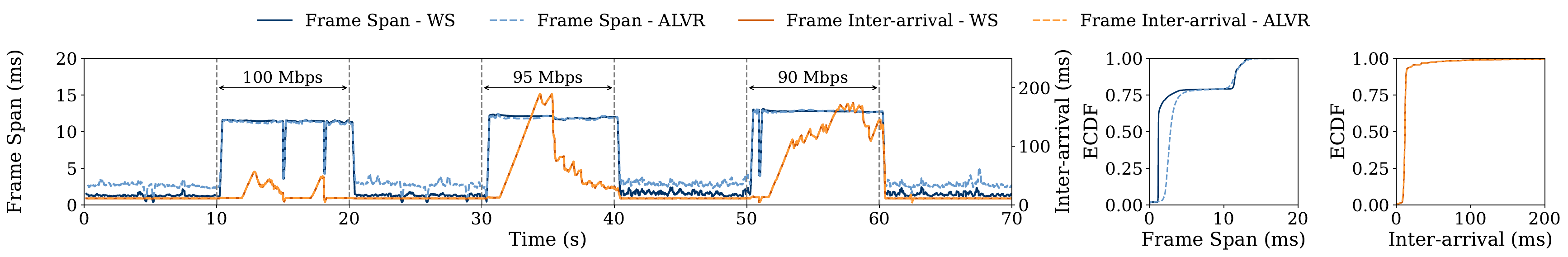}
    \vspace{-0.7cm} 

    \caption{}
    \label{fig:frame_span}
  \end{subfigure}
  
  \begin{subfigure}[b]{1.0\linewidth} 
    \includegraphics[width=\linewidth]{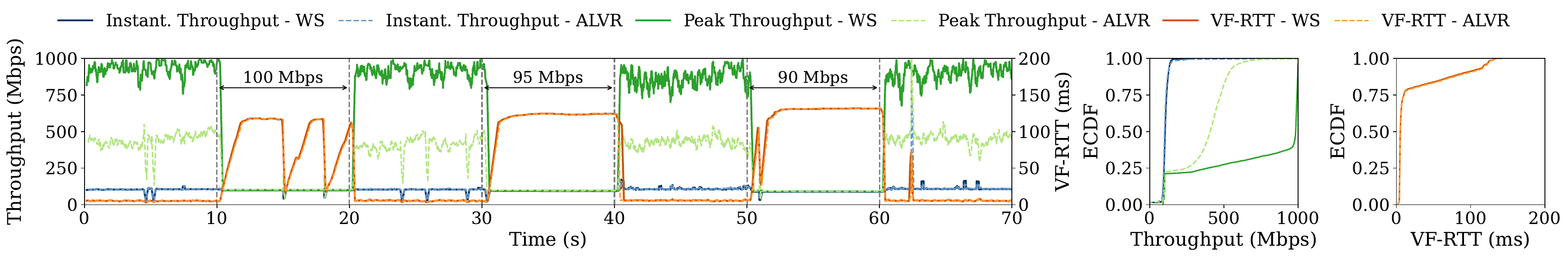}
    \vspace{-0.7cm} 
    \caption{}

    \label{fig:throughput}
  \end{subfigure}

  \begin{subfigure}[b]{1.0\linewidth} 
    \includegraphics[width=\linewidth]{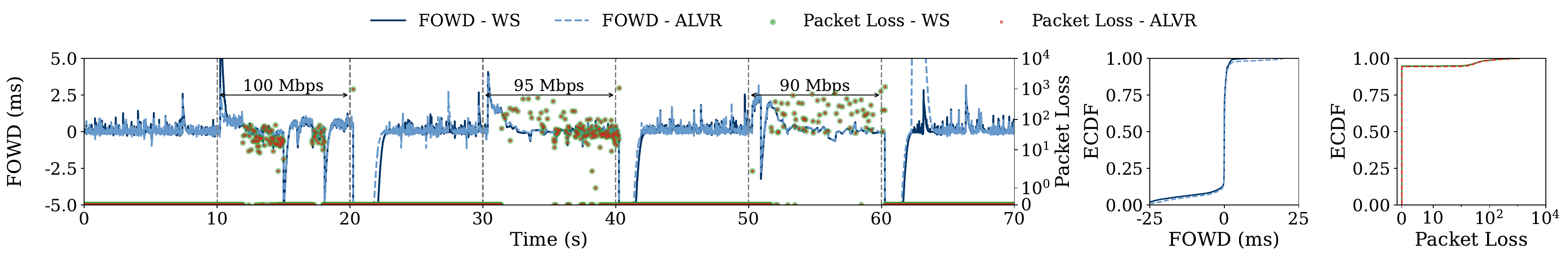}
    \vspace{-0.7cm} 

    \caption{}
    \label{fig:owdg}
  \end{subfigure}

  \vspace{-0.2cm} 

  \caption{Temporal evolution and Empirical Cumulative Distribution Functions (ECDF) of our metrics under a bandwidth-limited test (see $1$ in Tbl.~\ref{table:emulation_tests}), comparing the values logged in ALVR with those independently derived from Wireshark (WS) traces.
  Temporal evolutions are filtered using a 16-sample sliding window average to enhance visibility, except for packet loss (discrete) and FOWD (implicitly filtered).}

  \label{fig:combined_plots}
\end{figure*}

Fig.~\ref{fig:combined_plots} demonstrates that both the metrics logged within ALVR and their measurements derived from Wireshark traces exhibit similar distributions. However, as illustrated in Figs.~\ref{fig:combined_plots}a and \ref{fig:combined_plots}b, both the client-side frame span and peak network throughput differ under baseline conditions. This disparity can be attributed to Wi-Fi’s lower data rate compared to Ethernet’s 1~Gbps link, highlighting Wi-Fi as the system’s bottleneck when the bandwidth limitation is not enforced. 

Fig.~\ref{fig:combined_plots}b illustrates that peak network throughput is an effective estimator of network capacity during bandwidth-limited periods. Concurrently, there are significant increases in client-side frame span and VF-RTT due to the traffic controller limiting the packet departure rate. Additionally, substantial packet losses occur as the first-in-first-out queue of the netem overflows, leading to increased frame inter-arrival times. During transition periods, there are noticeable peaks in FOWD as the netem’s packet departure rate to the AP varies.

\section{NeSt-VR: Performance Assessment} \label{Sec:abr_evalu}

In this section, we compare CBR, ALVR’s ABR algorithm, and NeSt-VR in terms of their operation, behavior, and performance during an emulated bandwidth-limited test (Section \ref{Sec:comparison_capacity}) and a real mobility test (Section \ref{Sec:comparison_mobility}). Tbl.~\ref{tab:heur-values_test} outlines the NeSt-VR parameter values\footnote{$n$ is set to 256 to filter transient disruptions and ensure a conservative algorithm response.
$\gamma$ equals 0.25 to strike a balance between exploring higher bitrates and maintaining stability. $\rho$ equals 0.95 to guarantee that the VF delivery rate remains close to the transmission rate. $\varsigma$ is set to 2.0, ensuring VFs are generally received before the subsequent VF transmission} used during the tests. To compare their operation, the temporal evolution of the target bitrate and several key Quality of Service (QoS) metrics ---average video network throughput\footnote{
Average video network throughput is computed using a time-weighted average of instantaneous throughput values rather than an arithmetical average given non-uniform intervals between samples}, frame delivery rate, packet loss, and VF-RTT--- are illustrated. Note that, during testing, the server consistently achieved a transmission frame rate of 90~fps.

    

\begin{table}[t]
    \centering
    \caption{NeSt-VR parameter values used during testing.}
    \small
    \begin{tabular}{@{}ll|ll@{}}
    \toprule
    \textbf{Parameter} & \textbf{Value} & \textbf{Parameter} & \textbf{Value} \\
    \midrule
    $\tau$ & 1 s & $\beta$ & 10~Mbps \\
    $n$ & 256 & $m$ & 0.90 \\
    $B_{\min}$ & 10~Mbps & $\gamma$ & 0.25 \\
    $B_{\max}$ & 100~Mbps & $\rho$ & 0.95 \\
    $B_0$ & 30~Mbps & $\varsigma$ & 2.0 \\
    \bottomrule 
    \end{tabular}
    \label{tab:heur-values_test}
\end{table}

\begin{figure*}[t!!]
  \centering
  \captionsetup[subfigure]{oneside,margin={1.5cm,0cm}} 
  \begin{subfigure}[b]{0.389\linewidth}
    \includegraphics[width=\linewidth]{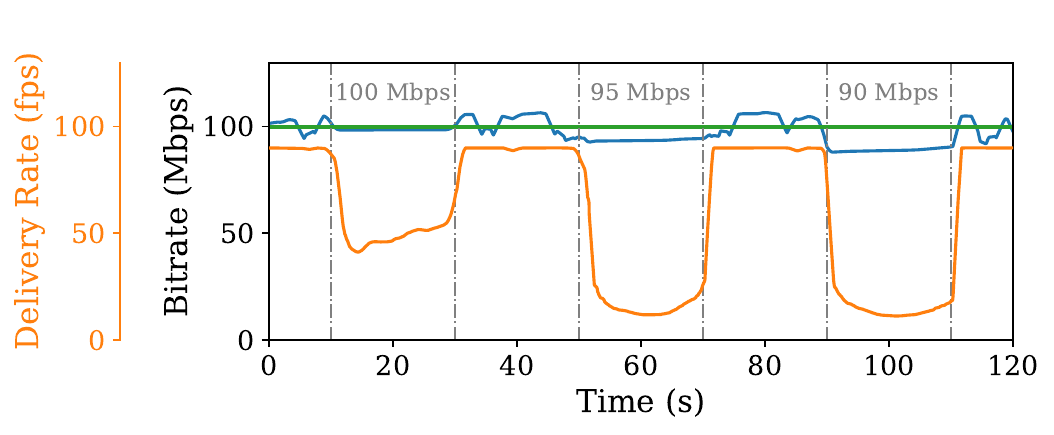}
  \end{subfigure}
  \captionsetup[subfigure]{oneside,margin={0cm,0cm}} 
  \begin{subfigure}[b]{0.299\linewidth}
    \includegraphics[width=\linewidth]{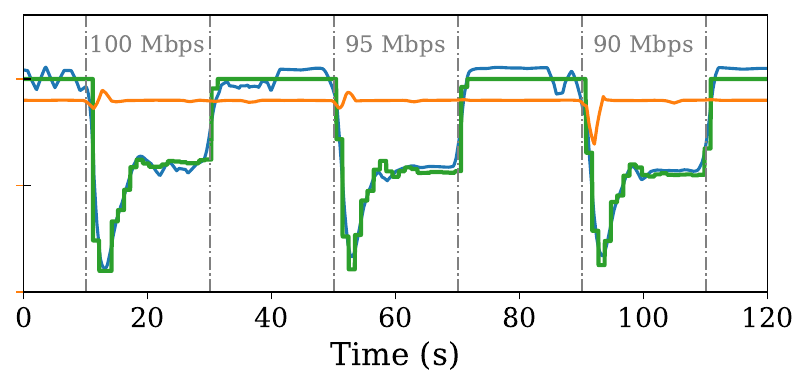}
  \end{subfigure}
  \begin{subfigure}[b]{0.299\linewidth}
    \includegraphics[width=\linewidth]{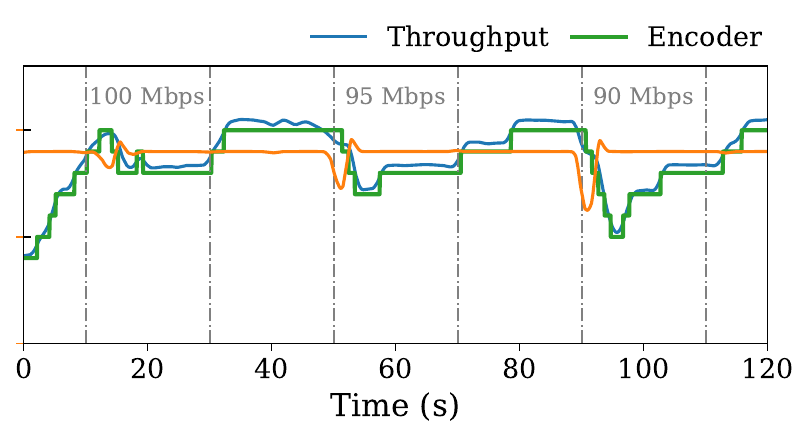}
  \end{subfigure}
  \captionsetup[subfigure]{oneside,margin={1.5cm,0cm}} 
  \begin{subfigure}[b]{0.389\linewidth}
    \includegraphics[width=\linewidth]{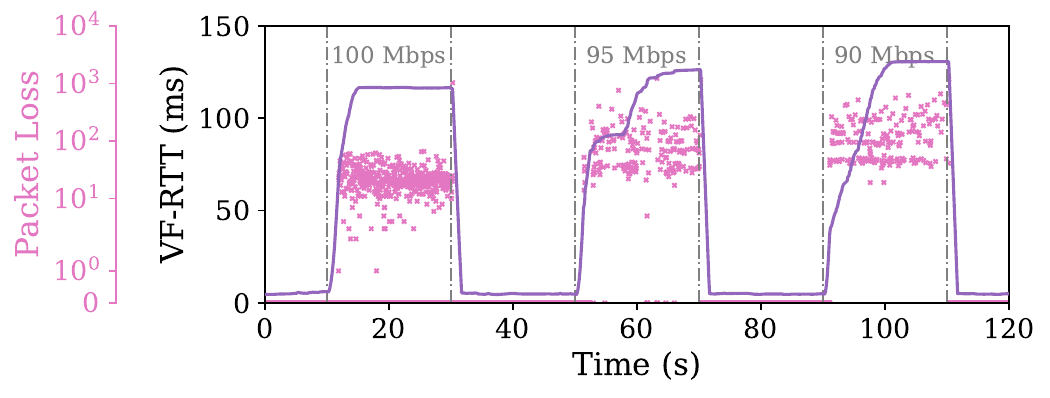}
    \caption{CBR}
  \end{subfigure}
  \captionsetup[subfigure]{oneside,margin={0cm,0cm}} 
  \begin{subfigure}[b]{0.299\linewidth}
    \includegraphics[width=\linewidth]{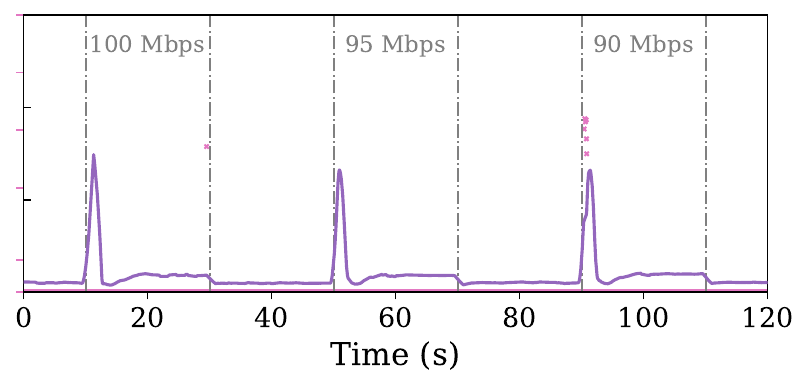}
    \caption{ALVR's ABR}
  \end{subfigure}
  \begin{subfigure}[b]{0.299\linewidth}
    \includegraphics[width=\linewidth]{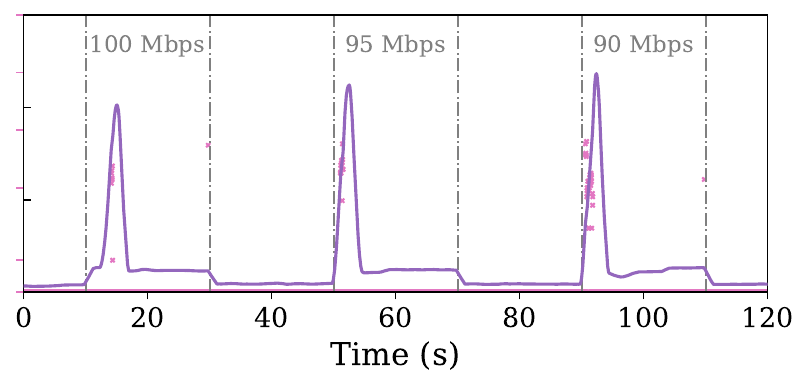}
    \caption{NeSt-VR}
  \end{subfigure}
  \caption{Comparison of CBR, ALVR's ABR, and NeSt-VR under emulated limited bandwidth conditions. Sliding window of 128 samples for frame delivery rate, average video network throughput, and VF-RTT.}
  \label{fig:comparison_bitrate_schemes}
\end{figure*}

\subsection{Network capacity fluctuations} \label{Sec:comparison_capacity}

During this test, the network capacity changes during three 20-second intervals. As in Section~\ref{Sec:validation}, the bandwidth is limited to the values outlined in Tbl.~\ref{table:emulation_tests} and the client is approximately placed 1 m from the AP, observing a -40~dBm RSSI. As depicted in Fig.~\ref{fig:comparison_bitrate_schemes}a, CBR maintains a constant 100~Mbps target bitrate regardless of the network's capacity, causing VF-RTTs to go above 90~ms and substantial packet losses during bandwidth-limited periods: 10,598, 18,346, and 26,277 packets lost during intervals 1, 2, and 3, respectively. This leads to significant reductions in frame delivery rate, averaging 55.2, 51.9, and 41.8~fps at each step.

In contrast, as illustrated in Fig.~\ref{fig:comparison_bitrate_schemes}b, ALVR's ABR rapidly decreases the target bitrate to the minimum (i.e., 10~Mbps) in response to increased delays stemming from constraining the network capacity. 
Then, since drastically reducing the bitrate leads to much smaller VFs, lowering network delay and VF-RTT, the algorithm starts increasing the target bitrate until stabilization around 60~Mbps, achieving an average target bitrate of 53.8, 54.4, and 52.5~Mbps at intervals 1, 2, and 3, respectively. 
Once the bandwidth limitation is deactivated, ALVR's ABR promptly increases the target to the maximum (i.e., 100~Mbps).
Thus, as illustrated, ALVR's ABR algorithm is able to adapt the bitrate when capacity changes, minimizing packet loss (717 losses in total) and maintaining high frame delivery rates (89.7~fps on average).

Similarly, as shown in Fig.~\ref{fig:comparison_bitrate_schemes}c, NeSt-VR is able to react to capacity changes, driven by a reduced frame delivery rate (leading to a NFR below $\rho$) and increased VF-RTTs (above $\sigma$~ms). However, instead of dropping the target bitrate to the minimum, it is progressively reduced in 10~Mbps steps. 
Upon removal of each limit in bandwidth, given that NFR is above $\rho$ and VF-RTT is below $\sigma$, NeSt-VR increases the target bitrate (with probability $\gamma$) in 10~Mbps steps, taking a more conservative approach. This behavior leads to higher target bitrates than ALVR's ABR algorithm (83.9, 79.5, and 72.4~Mbps on average at each step), while also sustaining a high frame delivery rate (89.4~fps on average) and reduced packet loss (1,088 packets lost in total).


\begin{figure*}[t!!]
  \centering
  \captionsetup[subfigure]{oneside,margin={1.5cm,0cm}} 
  \begin{subfigure}[b]{0.389\linewidth}
    \includegraphics[width=\linewidth]{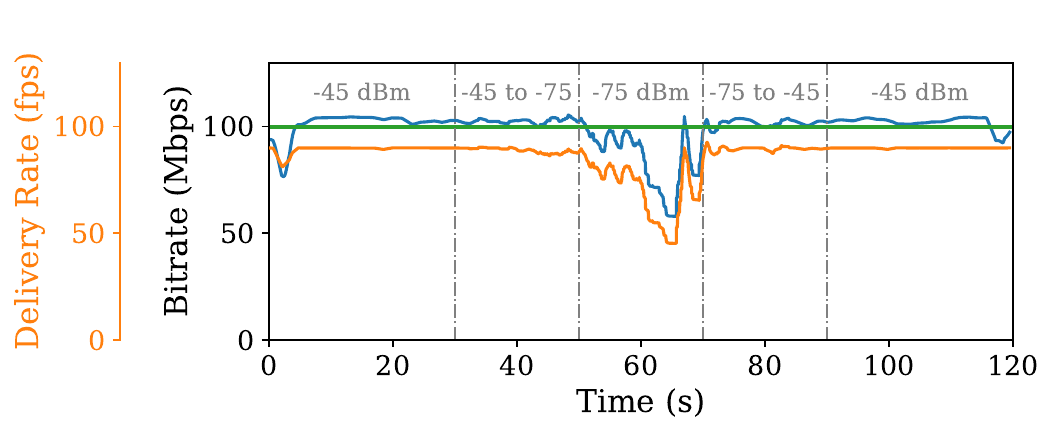}
  \end{subfigure}
  \captionsetup[subfigure]{oneside,margin={0cm,0cm}} 
  \begin{subfigure}[b]{0.299\linewidth}
    \includegraphics[width=\linewidth]{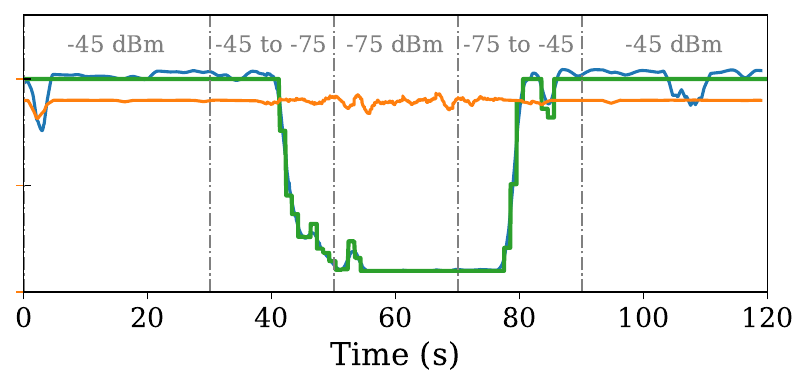}
  \end{subfigure}
  \begin{subfigure}[b]{0.299\linewidth}
    \includegraphics[width=\linewidth]{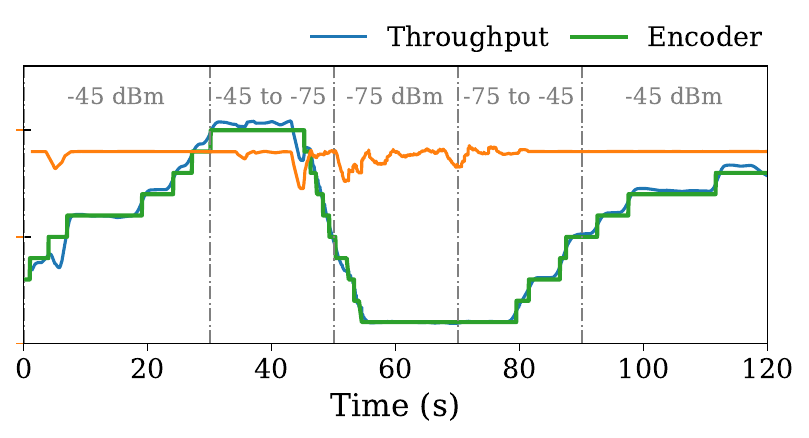}
  \end{subfigure}
  \captionsetup[subfigure]{oneside,margin={1.5cm,0cm}} 
  \begin{subfigure}[b]{0.389\linewidth}
    \includegraphics[width=\linewidth]{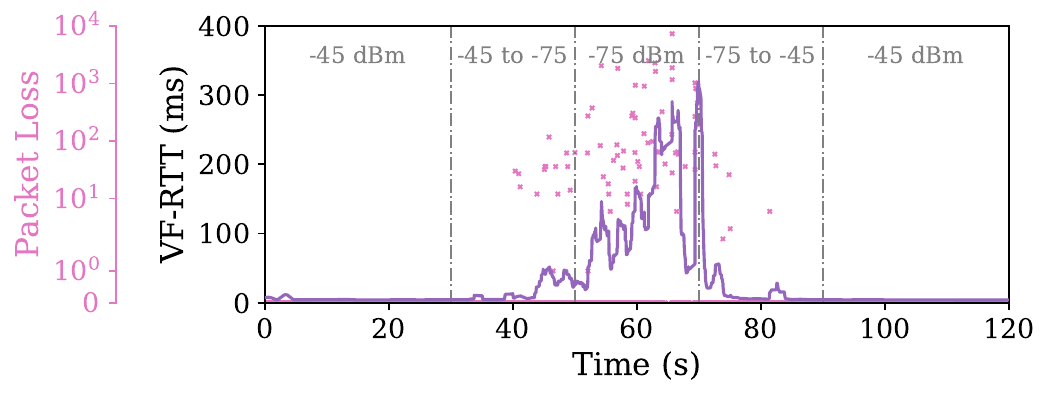}
    \caption{CBR}
  \end{subfigure}
  \captionsetup[subfigure]{oneside,margin={0cm,0cm}} 
  \begin{subfigure}[b]{0.299\linewidth}
    \includegraphics[width=\linewidth]{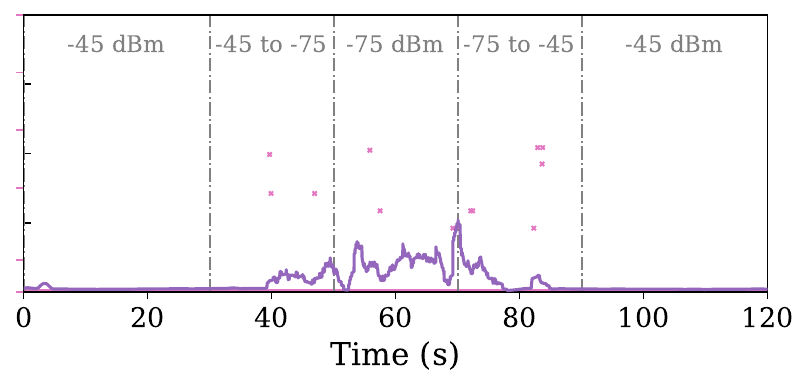}
    \caption{ALVR's ABR}
  \end{subfigure}
  \begin{subfigure}[b]{0.299\linewidth}
    \includegraphics[width=\linewidth]{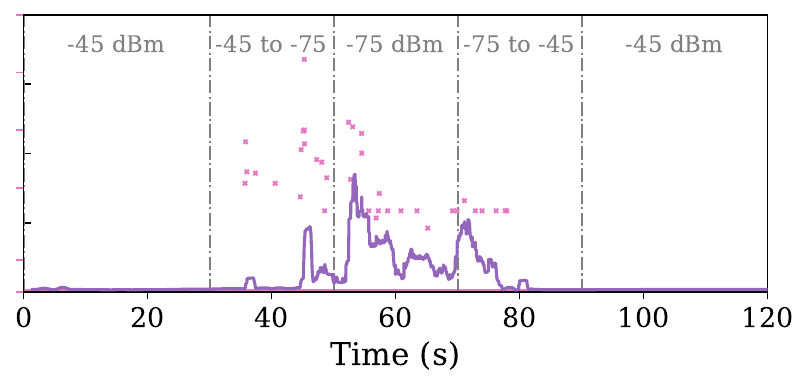}
    \caption{NeSt-VR}
  \end{subfigure}
  \caption{Comparison of CBR, ALVR's ABR, and NeSt-VR during a mobility test. Sliding window of 128 samples for frame delivery rate, average video network throughput, and VF-RTT.}
  \label{fig:comparison_bitrate_schemes_mobility}
\end{figure*}

\subsection{Mobility} \label{Sec:comparison_mobility}

During this mobility test, a user temporarily moves to a distant point with an RSSI of approximately -75~dBm, remaining there for 20 seconds before returning. 
As illustrated in Fig.~\ref{fig:comparison_bitrate_schemes_mobility}a, using CBR, the significant decrease in RSSI as the user moves away from the AP ---and the subsequent reduction in Wi-Fi transmission rates--- results in multiple packet losses, leading to substantial drops in both throughput and frame delivery rate. For instance, at the farthest point (i.e., -75~dBm), the VF-RTT averages 141.2~ms, with 29,421 packets lost, significantly reducing the frame delivery rate to 74.7~fps on average.
In contrast, Fig.~\ref{fig:comparison_bitrate_schemes_mobility}b demonstrates that ALVR's ABR rapidly responds to increased network delays, lowering the bitrate to the minimum. At the farthest point, the target bitrate averages 11.2~Mbps, significantly reducing VF-RTT compared to CBR (42.8~ms on average) while maintaining a high frame delivery rate (89.3~fps on average) due to minimal packet loss (53 packets lost in total). Similarly, as the user moves back closer to the AP and RSSI improves, ABR promptly increases the target bitrate to the maximum.
Likewise, Fig.~\ref{fig:comparison_bitrate_schemes_mobility}c shows that NeSt-VR also decreases the bitrate significantly when RSSI worsens, but in a more controlled manner. At the farthest point, the target bitrate averages 14.6~Mbps, with a moderate VF-RTT (60.9 ms on average) and a frame delivery rate averaging 87.2~fps, given 436 packets lost. As RSSI improves, NeSt-VR progressively increments the bitrate conservatively to maintain a more consistent image quality.

\section{Conclusions}

In this work, we have presented and validated a set of metrics integrated into ALVR. These metrics provide essential insights into the state and performance of the network, aiding bitrate adaptation algorithms in their decision-making process. Indeed, we have designed and implemented NeSt-VR, an ABR algorithm that uses several of these metrics to dynamically adjust the bitrate based on the network performance. In particular, NeSt-VR reduces the bitrate during network congestion intervals, characterized by unsuccessful VF deliveries or prolonged delivery times. 


In future research, we aim to design and test more sophisticated bitrate adaptation algorithms tailored to single and multi-user VR scenarios within the presented ALVR framework. These ABR algorithms will incorporate users' QoE indicators, focusing on aspects such as image quality, immersion, and interactivity. 




\section{Acknowledgement}

This work is partially funded by MAX-R (101070072) EU, Wi-XR PID2021-123995NB-I00 (MCIU/AEI/FEDER,UE), and by MCIN/AEI under the Maria de Maeztu Units of Excellence Programme (CEX2021-001195-M). 

\bibliographystyle{unsrt} 
\bibliography{bib}

\begin{thebibliography}{10}

\bibitem{minopoulos2022opportunities}
Georgios Minopoulos and Konstantinos~E Psannis.
\newblock {Opportunities and challenges of tangible XR applications for 5G networks and beyond}.
\newblock {\em IEEE Consumer Electronics Magazine}, 12(6):9--19, 2022.

\bibitem{akyildiz2022wireless}
Ian~F Akyildiz and Hongzhi Guo.
\newblock {Wireless extended reality (XR): Challenges and new research directions}.
\newblock {\em ITU J. Future Evol. Technol}, 3(2):1--15, 2022.

\bibitem{wifialliance_VR_reqs}
{Wi-Fi Alliance}.
\newblock {{Wi-Fi Delivers Immersive VR Gaming}}.
\newblock Technical report, Wi-Fi Alliance, 2023.

\bibitem{giordano2023will}
Lorenzo~Galati Giordano, Giovanni Geraci, Marc Carrascosa, and Boris Bellalta.
\newblock {What will Wi-Fi 8 be? A primer on IEEE 802.11 bn ultra high reliability}.
\newblock {\em arXiv preprint arXiv:2303.10442}, 2023.

\bibitem{ABR_survey}
Yusuf Sani, Andreas Mauthe, and Christopher Edwards.
\newblock {Adaptive Bitrate Selection: A Survey}.
\newblock {\em IEEE Communications Surveys \& Tutorials}, 19(4):2985--3014, 2017.

\bibitem{buffer-based-netflix}
Te-Yuan Huang, Ramesh Johari, Nick McKeown, Matthew Trunnell, and Mark Watson.
\newblock {A buffer-based approach to rate adaptation: evidence from a large video streaming service}.
\newblock {\em SIGCOMM Comput. Commun. Rev.}, 44(4):187–198, aug 2014.

\bibitem{BOLA}
Kevin Spiteri, Rahul Urgaonkar, and Ramesh~K. Sitaraman.
\newblock {BOLA: Near-Optimal Bitrate Adaptation for Online Videos}.
\newblock {\em IEEE/ACM Transactions on Networking}, 28(4):1698--1711, 2020.

\bibitem{liubogoshchev2021everest}
{Liubogoshchev, Mikhail and Korneev, Evgeny and Khorov, Evgeny}.
\newblock Everest: Bitrate adaptation for cloud vr.
\newblock {\em Electronics}, 10(6):678, 2021.

\bibitem{korneev2024model}
Eugene Korneev, Mikhail Liubogoshchev, Dmitry Bankov, and Evgeny Khorov.
\newblock {How to Model Cloud VR: An Empirical Study of Features That Matter}.
\newblock {\em IEEE Open Journal of the Communications Society}, 2024.

\bibitem{ALVR_5G_DQN}
Yaohua Sun, Jianmin Chen, Zeyu Wang, Mugen Peng, and Shiwen Mao.
\newblock {Enabling Mobile Virtual Reality with Open 5G, Fog Computing and Reinforcement Learning}.
\newblock {\em IEEE Network}, 36(6):142--149, 2022.

\bibitem{GCC-webrtc}
Gaetano Carlucci, Luca De~Cicco, Stefan Holmer, and Saverio Mascolo.
\newblock {Analysis and design of the google congestion control for web real-time communication (WebRTC)}.
\newblock In {\em Proceedings of the 7th International Conference on Multimedia Systems}, MMSys '16, New York, NY, USA, 2016. Association for Computing Machinery.

\bibitem{alhilal2024fovoptix}
Ahmad Alhilal, Ze~Wu, Yuk~Hang Tsui, and Pan Hui.
\newblock {FovOptix: Human Vision-Compatible Video Encoding and Adaptive Streaming in VR Cloud Gaming}.
\newblock In {\em Proceedings of the 15th ACM Multimedia Systems Conference}, pages 67--77, 2024.

\bibitem{vergados2023adaptive}
Dimitrios~J Vergados, Angelos Michalas, Alexandros-Apostolos~A Boulogeorgos, Spyridon Nikolaou, Nikolaos Asimopoulos, and Dimitrios~D Vergados.
\newblock {Adaptive Virtual Reality Streaming: A Case for TCP}.
\newblock {\em IEEE Transactions on Network and Service Management}, 2023.

\bibitem{rfc3550}
Henning Schulzrinne, Stephen~L. Casner, Ron Frederick, and Van Jacobson.
\newblock {RTP: A Transport Protocol for Real-Time Applications}.
\newblock RFC 3550, July 2003.

\end{thebibliography}

\end{document}